\documentclass[aps,pra,showpacs,twocolumn,superscriptaddress]{revtex4-1}
\usepackage{amsmath}
\usepackage{amsfonts}
\usepackage{amssymb}
\usepackage{graphicx}
\usepackage{subfig}
\pdfoutput=1

\begin{document}
\title{Trace distance and scaling behavior of coupled cavity lattice at finite temperature}
\author{Da-Wei Luo}
\author{Jing-Bo Xu}
\email{xujb@zju.edu.cn}
\affiliation{Zhejiang Institute of Modern Physics and Department of Physics, Zhejiang University, Hangzhou 310027, P.R.China}
\date{\today}

\begin{abstract}
We use an alternative approach to study the quantum phase transition in a coupled cavity lattice at finite temperature. As an illustrative example, we investigate the behaviors of the trace distance and quantum phase transition in a Jaynes-Cummings lattice at finite temperature. It is found that the trace distance can be used to describe the critical point of the quantum phase transition at finite low temperatures and the critical points are sensitive to the atom-field interaction strength and the detuning factor. For non-equilibrium states, we demonstrate that the time evolution of the trace distance's maximum value is also a good indicator of the critical points. Moreover, we show that the scaling behavior of derivative of the trace distance at the critical points and the scaling rule are dependent on the external parameters of the Hamiltonian.
\end{abstract}
\pacs{64.60.-i, 42.50.Pq}
%General studies of phase transitions, Cavity quantum electrodynamics,

\maketitle

\section{Introduction}

Quantum phase transition (QPT) has been a hot topic in condensed matter physics over the years~\cite{Osborne2002,Sachdev1999}. The existence of a QPT strongly influences the behavior of many-body systems near the critical point associated with the divergence of correlation length of two-point correlation functions and the vanishing of the gap in the exciton spectrum. QPTs, which happen at very low temperature and are driven by pure quantum fluctuations, are a qualitative change in the ground state properties of a quantum many-body system as some external parameters of the Hamiltonian are varied.

Recently, the Jaynes-Cummings-Hubbard (JCH) lattice~\cite{Greentree2006} has been shown to display the quantum phase transition phenomenon~\cite{Illuminati2006, Hartmann2008,Quach2009,Greentree2006,Makin2008,Tan2011} within the mean-field theory framework and verified by Monte Carlo simulations~\cite{Pippan2009,Aichhorn2008}. For large numbers of coupled cavity QED systems, it should be possible to observe many-body effects such as quantum phase transition. A strong coupling theory for the JCH lattice has also been developed~\cite{Schmidt2009}. The QPT of JCH lattice is analogous to the insulator-superfluid transition of the Bose-Hubbard model which has been theoretically and experimentally demonstrated to be realizable in cold-atom optical lattices~\cite{Greiner2002,Jaksch1998}. The advantage of coupled cavity system is that each lattice can be easily addressed and the system parameters can be readily controlled because of their mesoscopic size. Experimentally, the JCH lattice is much easier to realize and more controllable than strongly correlated systems at the level of each individual elements, and is still able to simulate the behavior of such systems~\cite{Illuminati2006}, and can be realized using superconducting circuits~\cite{Houck2012}.

On the other hand, the trace distance~\cite{Nielsen2000,Gilchrist2005} has been shown to serve as a measure for the distinguishability of quantum states as well as the non-Markovianity of quantum processes~\cite{Breuer2009,Laine2010a} and the witness for initial system-environment correlations in open-system dynamics~\cite{Laine2010,Smirne2010}. Moreover, the trace distance can be experimentally obtained using technologies such as quantum state tomography~\cite{Nielsen2000}. The trace distance between any two states is a direct measure of how far apart the two states are in the state space. Therefore, the trace distance between a state and its factorized state defined as the tensor product of the system state and the environment state can serve as a measure of in-separateness or the correlation between the environment and system~\cite{Smirne2010}. This approach, which compares a state and its factorized state is different from the fidelity approach which compares two ground states whose Hamiltonian parameters are slightly varied. In this paper we extend the ground states to Gibbs states and investigate the behavior of the trace distance and QPT of the JCH lattice at finite temperatures. The system is defined to be all the atoms and the environment to be all the field modes in our coupled Jaynes-Cumming lattice. Therefore, the trace distance between finite-temperature Gibbs initial equilibrium state and the factorized state defined as the tensor product of the system state and the environment state is a measure of the correlation between all the atoms and all the field modes. At the critical points of the quantum phase transition, the structure of the ground state undergoes a radical change. Just like the fidelity approach captures this radical change by calculating the inner product of two ground or Gibbs states whose  Hamiltonian parameters are slightly varied, we expect the trace distance measure to be able to pinpoint the critical points of the quantum phase transition, and our approach is better suited for interacting atom-field systems. By making use of the analytical solution to the JCH lattice of arbitrary size $N$, we calculate the trace distance between the Gibbs state and the product of its marginal states. At the critical points of the quantum phase transition, which are determined by the ground state energy level crossing, the trace distance shows a sudden jump at finite temperature, which means that the trace distance can be used to describe the critical points of QPT. The critical points are found to be dependent on the atom-field interaction strength and the detuning factor, from which we can obtain a phase diagram of the system. Moreover, non-equilibrium states are also taken into consideration, and it is found that the time evolution of the trace distance's maximum value is also a good indicator of the critical points. Finally, the scaling behavior of the system is found to exist for the first derivative of the trace distance at the critical points, and the scaling rule is shown to be dependent on the system parameters. This paper is organized as follows. In Section~\ref{sec_sys} we give the energy spectrum and phase diagram of the JCH lattice. The trace distance and the QPT behavior of the JCH lattice is explored in Section~\ref{sec_qpt}. The trace distance of the Gibbs equilibrium states for non-equilibrium time evolution is also studied. In Section~\ref{sec_scal}, the scaling behavior of the JCH lattice is investigated. A conclusion of the paper is given in Section~\ref{sec_con}.

\section{Energy spectrum and phase diagam of the JCH lattice}\label{sec_sys}

\begin{figure}
  \includegraphics[scale=.5]{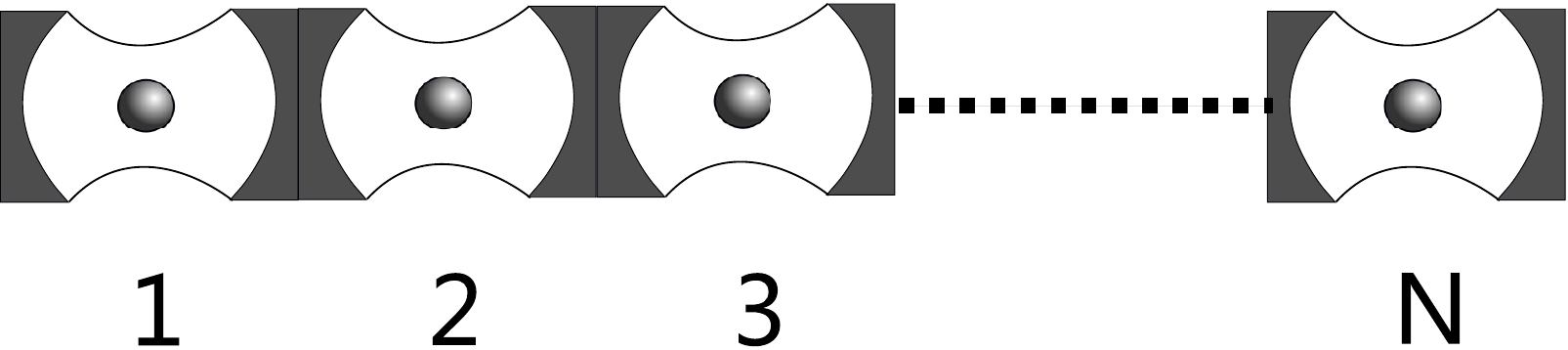}\\
  \caption{This is the schematic diagram of the system studied in this paper. Each cavity is modeled by the Jaynes-Cummings Hamiltonian and only nearest-neighbor hopping of photon is allowed.}\label{jchmod}
\end{figure}

The JCH lattice describes a system of $N$ low-loss cavities that are coupled together which allows nearest-neighbor photon hopping. Each low-loss cavity is modeled by the Jaynes-Cummings Hamiltonian~\cite{Jaynes1963} which sustains a single mode field $\omega_f$ and contains a two-level atom of Bohr frequency $\omega_a$ which couples to the field mode at rate $g$ (See Fig.~\ref{jchmod}). The full Hamiltonian reads
\begin{equation*}
H=\sum_{j=1}^N H_{JC}^{j}+H_{int},
\end{equation*}
where $H_{JC}^{j}$ is the Hamiltonian describing the $j$-th JC cavity of the form
\begin{equation}
	H_{JC}^{j}=\omega_a \sigma^+_j \sigma^-_j+\omega_f b_j^\dagger b_j + g (\sigma^+_j b_j+\sigma^-_j b_j^\dagger),
\end{equation}
and $\sigma^+=|1 \rangle\langle 0|$ and $\sigma^-=|0 \rangle\langle 1|$ are the raising and lowering operators of the atom, $b^\dagger$ and $b$ are the creation and annihilation operators of the field mode and the interaction Hamiltonian is of the form $H_{int} = -\kappa (b_{j+1}^\dagger b_j + h.c.)$. The Hamiltonian can be written in a decoupled form using the Fourier transform. We can express the free-field Hamiltonian $H_{free}=\omega_f b_j^\dagger b_j -\kappa (b_{j+1}^\dagger b_j + h.c.) $ in terms of normal modes as~\cite{Ciccarello2011,Makin2009,Ogden2008}
\begin{equation}
H_{free}=\sum_k \omega_k \alpha_k ^\dagger \alpha_k,
\end{equation}
where
\begin{eqnarray*}
k&=&\frac{2\pi m}{N+1} \quad (m=1,\ldots, N),\\
\omega_k&=&\omega_f+2\kappa\cos\frac{k}{2},\\
\alpha_k&=&\sqrt{\frac{2}{N+1}}\sum_{i=1}^N\sin\left(\frac{k}{2}i\right)b_i.
\end{eqnarray*}
Since the atom-photon interaction strengths, the cavity-mode and atomic frequencies are uniform through the cavity lattices, we can rewrite the Hamiltonian in terms of $N$ decoupled effective JC models as
\begin{equation}
H(\omega_k,\omega_a,g)=\sum_k\left[\omega_k\alpha_k ^\dagger \alpha_k +\omega_a s_k ^\dagger s_k+g \left(\alpha_k^\dagger s_k+s_k^\dagger\alpha_k\right)\right],
\end{equation}
where $s_k$ and $s_k ^\dagger$ are the atomic lowering and raising operators after a similar transform. The Hamiltonian now takes the form of $N$ uncoupled JC cavities with different, specific field mode frequencies. For the $k$-th effective JC cavity, the eigenvectors are given by, in the basis $\{|\mathrm{atom,field}\rangle_k\}$,
\begin{align}
|\varphi_n^+(k)\rangle&=a_n(k)|1,n-1\rangle_k+b_n(k)|0,n\rangle_k,&\nonumber\\
|\varphi_n^-(k)\rangle&=-b_n(k)|1,n-1\rangle_k+a_n(k)|0,n\rangle_k,&\nonumber\\
|\varphi_0^-(k)\rangle&=|0,0\rangle_k,&\nonumber
\end{align}
with
\begin{equation}
a_n(k)=\sqrt{\frac{\Omega_n(k)+\Delta_k}{2\Omega_n(k)}},\quad b_n(k)=\sqrt{\frac{\Omega_n(k)-\Delta_k}{2\Omega_n(k)}},
\end{equation}
where $\Omega_n(k)=\sqrt{\Delta_k^2+4g^2n}$ and $\Delta_k=\omega_a-\omega_k$ is the detuning of the $k$-th effective cavity. The corresponding energy levels are given by
\begin{equation}
E_n^\pm(k)=n \omega_k+\frac{\Delta_k}{2}\pm\frac{\Omega_n(k)}{2},\quad E_0^-(k)=0.
\end{equation}

\begin{figure}
  \includegraphics[scale=.6]{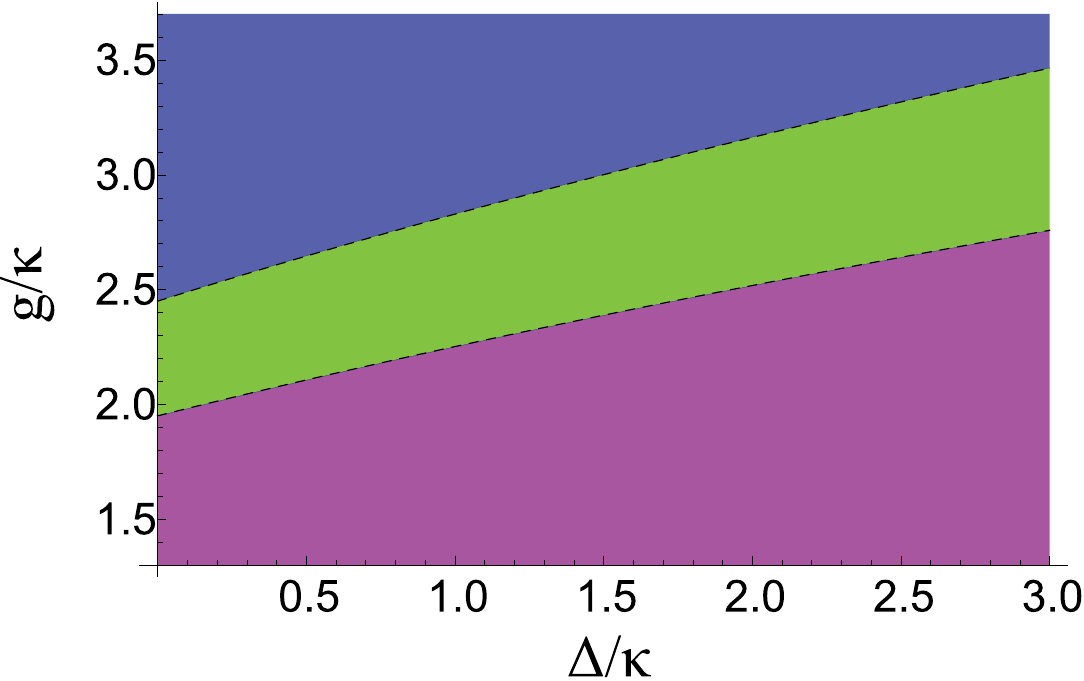}\\
  \caption{(Color online) Phase diagram of the JCH lattice. At absolute zero temperature, the ground state level crossing happens whenever the condition in Eq.~\eqref{gnd_cr_g} is met, which is defined to be the critical points of QPT. In the phase diagram, the critical points lie on the dashed black line, and each phase has a distinctive ground state wave function associated with it.}\label{qpt_dg}
\end{figure}

\begin{figure}
  \includegraphics[scale=.75]{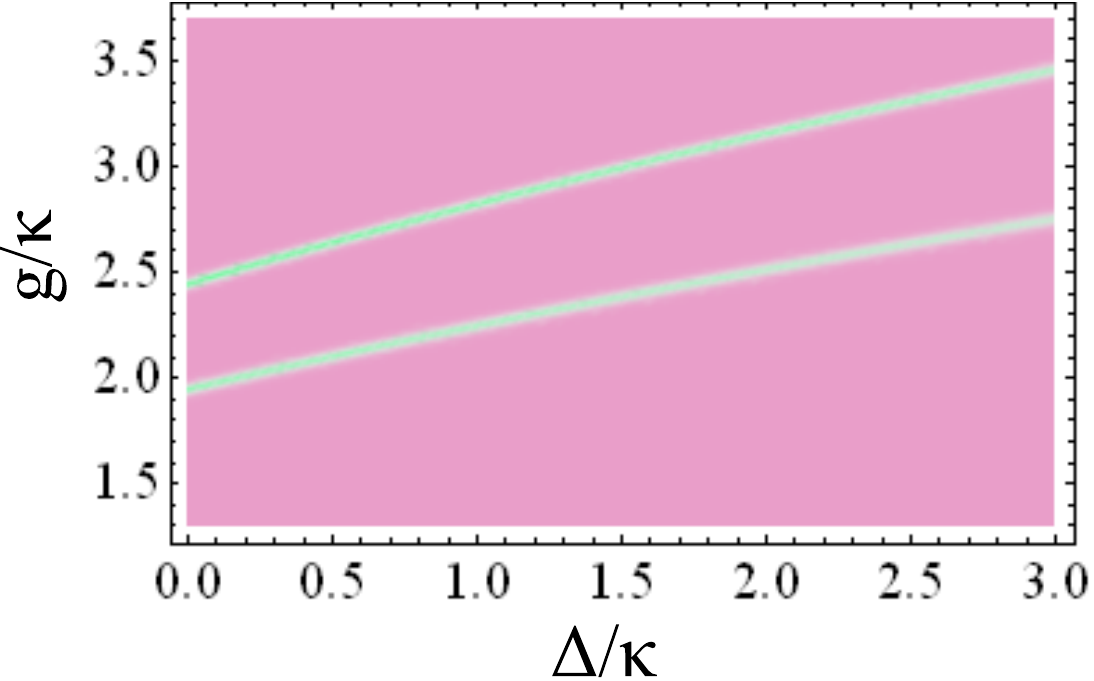}\\
  \caption{(Color online) Fidelity of two Gibbs states near zero temperature with slightly varied coupling $\delta g=0.01$ as a function of the coupling strength $g$ and detuning $\Delta$. It can be readily seen that along the critical points, the fidelity displays a sudden drop, which agrees exactly with the phase diagram.}\label{fid0}
\end{figure}

\begin{figure}
  \includegraphics[scale=.6]{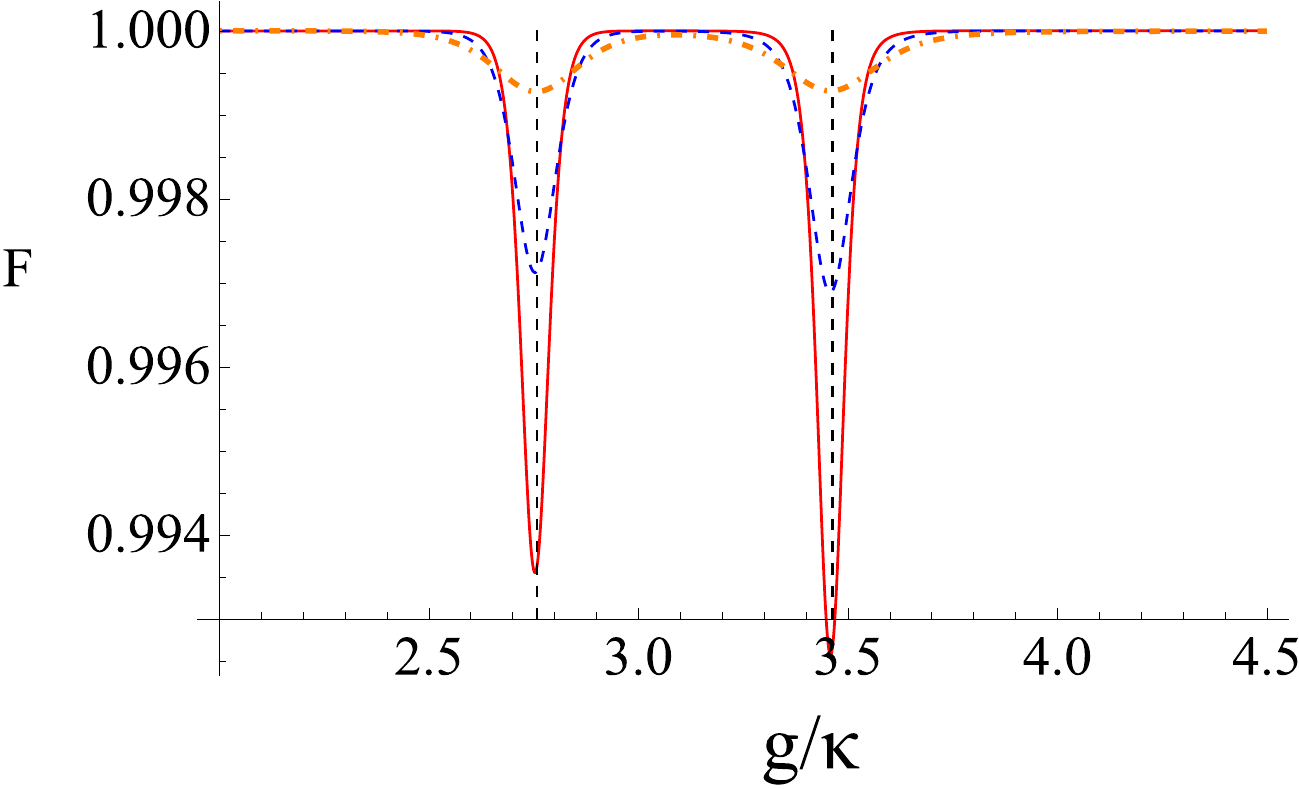}\\
  \caption{(Color online)  Fidelity of two Gibbs states at various temperatures with slightly varied coupling $\delta g=0.01$ as a function of the coupling strength $g$. The inverse temperature is taken to be $\beta=20$ (Orange dash-dotted line), $\beta=40$ (Blue dashed line) and $\beta=60$ (Red solid line), and the vertical black dashed lines signify the critical points. We can see from the figure that the drop of fidelity is less dramatic for higher temperatures.}\label{fidt}
\end{figure}

Each ground state level crossing happens when two smallest eigen energies $E_n^-(k)$ coincide. In Fig.~\ref{qpt_dg}, we display the phase diagram of the JCH lattice at zero temperature. For simplicity, we choose the parameters $\omega_f/\kappa=3$ and $\Delta_f/\kappa\in[0,3]$ and $N=5$. The ground state crossing of three lowest-lying energy levels happens when
\begin{align}
g_c^{(1)}/\kappa&=\sqrt{\frac{\omega_f ^2}{\kappa^2}+\frac{\omega_f}{\kappa} \frac{\Delta_f}{\kappa} +2(\frac{\Delta_f}{\kappa} +\frac{\omega_f}{\kappa} )\cos\left[\frac{5\pi }{6}\right]},\nonumber\\
\mathrm{or,}&&\nonumber\\
g_c^{(2)}/\kappa&=\sqrt{\frac{\omega_f ^2}{\kappa^2}+\frac{\omega_f}{\kappa} \frac{\Delta_f}{\kappa} +2(\frac{\Delta_f}{\kappa} +\frac{\omega_f}{\kappa} )\cos\left[\frac{4\pi }{6}\right]},\label{gnd_cr_g}
\end{align}
which are the critical points of the quantum phase transition, and they divide the parameter space into three parts, which we plot as the dashed black lines in the phase diagram Fig.~\ref{qpt_dg}. One of the advantages of our approach is that we do not require prior knowledge of the order parameters or the pattern of the symmetry breaking to study the quantum phase transition phenomenon. The different phases are represented in terms of different ground states, each region in the phase diagram has its own ground state wave function, and the quantum phase transition happens when the system parameter changes from one region to another. Because of this ground state cross over, it is expected that the structure of the ground state undergoes a radical change when the quantum phase transition takes place. In order to compare our results with the fidelity approach, we explore the inner product or fidelity~\cite{GU2010} of two ground states with slightly varied parameters and find that it suffers a sudden drop at the critical points of the QPT. We also calculate the fidelity at finite temperature. The ground state wave function is then replace by the Gibbs equilibrium state density operator, which reduces to the ground state at absolute zero temperature. The fidelity of any two density operators $\rho$, $\sigma$ is given by $F(\rho,\sigma)=\mathrm{Tr}[\sqrt{\sigma^{1/2}\rho\sigma^{1/2}}]$. The Gibbs state of the JCH lattice can be written as a function of the coupling strength $g$. Taking
\begin{align*}
&\rho=\exp[-\beta H(\omega_k,\omega_a,g)]/Z_\rho, \\
&\mathrm{where}\quad Z_\rho=\mathrm{Tr}[\exp[-\beta H(\omega_k,\omega_a,g)]]; \\
&\sigma=\exp[-\beta H(\omega_k,\omega_a,g+\delta g)]/Z_\sigma, \\
&\mathrm{where}\quad Z_\sigma=\mathrm{Tr}[\exp[-\beta H(\omega_k,\omega_a,g+\delta g)]],
\end{align*}
where the inverse temperature $\beta=1/k_b T$ and $k_b$ the Boltzmann constant, we plot the fidelity in Fig.~\ref{fid0} at near zero temperature with $\beta=100$ and $\delta g=0.01$. It can be readily seen that along the critical points, the fidelity displays a sudden drop, which agrees exactly with the phase diagram we obtained. The Gibbs states fidelity also has a strong dependence on temperature. Choosing the inverse temperatures $\beta=20, 40, 60$, we plot the fidelity in Fig.~\ref{fidt}. We can see from the figure that the drop of fidelity is less dramatic for higher temperatures, which is due to thermal fluctuations at higher temperatures.

\begin{figure}
  \centering
  \subfloat[]{\includegraphics[scale=.55]{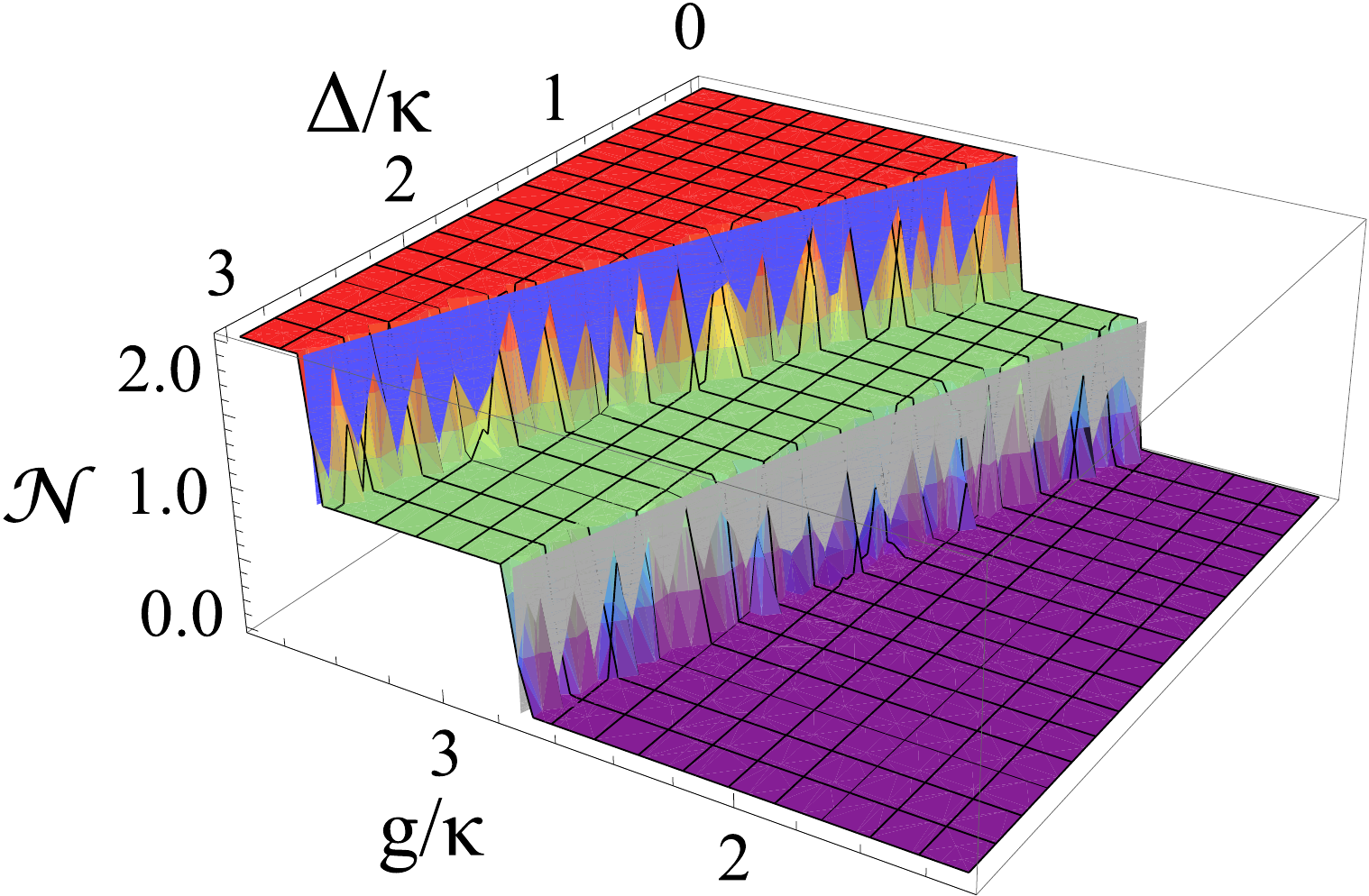}}\\
  \subfloat[]{\includegraphics[scale=.6]{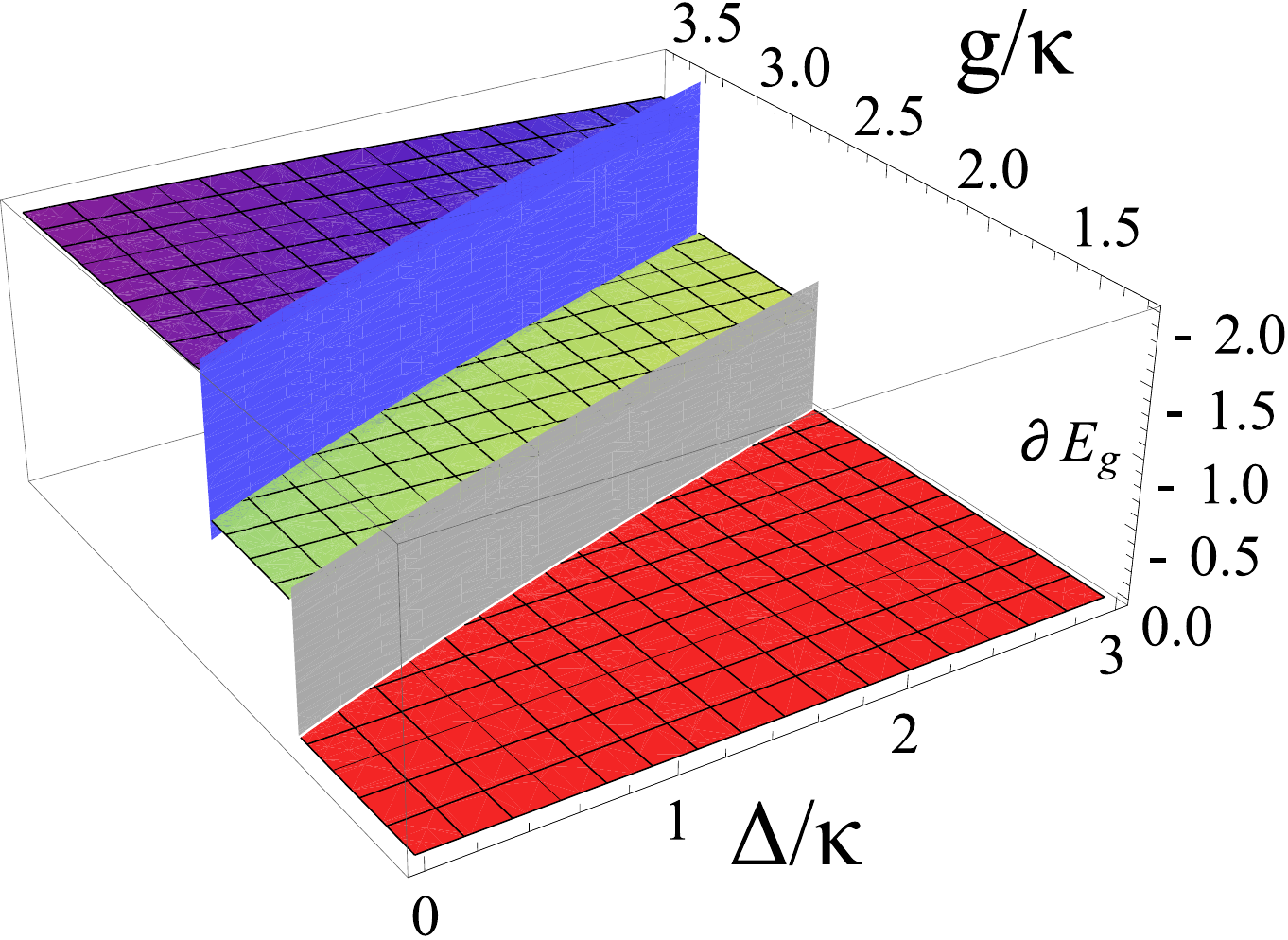}}\\
  \caption{(Color online) Expectation value of total excitation $N$ at finite low temperature as a function of the atom-field coupling strength $g$ and the detuning $\Delta$ with $\kappa\beta=800$(panel (a)), and the derivative of the ground state $E_g$ with critical points interposed(panel (b)). The vertical blue and gray surfaces indicate the critical points. It can be seen that at the critical points, both display a sudden change in value.}\label{qpt_en}
\end{figure}

As different phases are represented in terms of different ground states, and each phase has a distinctive ground state wave function associated with it, it is expected that the expectation value of the total excitation number, the derivative of the ground state energy as well as the trace distance should show a discontinuity behavior along the critical points of the quantum phase transition, even at finite temperatures. To verify this, we first plot the expectation value of total excitation $\mathcal{N}=\sum_k\langle s_k^+s_k^-+\alpha_k^\dagger\alpha_k\rangle$ as a function of the atom-field coupling strength $g$ and the detuning $\Delta$ near zero temperature $\kappa\beta=800$ and the derivative of the ground state $E_g$ against the atom-field coupling strength $g$ in Fig.~\ref{qpt_en}. It is quite clear that as a result of the structural change of the ground state in each phase, the total excitation $\mathcal{N}$ and derivative of the ground state energy has a different value and shows a sudden jump along the critical points defined by Eq.~\eqref{gnd_cr_g}, which indicates that there exists a QPT along the critical points.

\section{Trace distance and the quantum phase transition of the JCH lattice}\label{sec_qpt}

In this section, we investigate the trace distance and how to identify the QPT of the JCH lattice using the trace distance. The trace distance has recently been shown to be able to witness initial system-environment correlations in open-system dynamics and to distinguish quantum states. The distance of two trace class operators $\rho_1$ and $\rho_2$ is defined to be half the trace norm of $\rho_1-\rho_2$. For density operators, the trace distance can be further simplified as
\begin{equation}
D(\rho_1,\rho_2)=\frac{1}{2}\sum_i|d_i|\label{dst_def0},
\end{equation}
where $d_i$ are the eigenvalues of $\rho_1-\rho_2$. The trace distance ranges from zero to one, with its being zero if and only if the two states are identical. It is also a metric on the space of physical states and is sub-additive with respect to the tensor product,
\begin{equation*}
D(\rho_a\otimes\rho_1,\rho_b\otimes\rho_2)\leq D(\rho_a,\rho_b)+D(\rho_1,\rho_2).
\end{equation*}

\begin{figure}
\includegraphics[scale=.5]{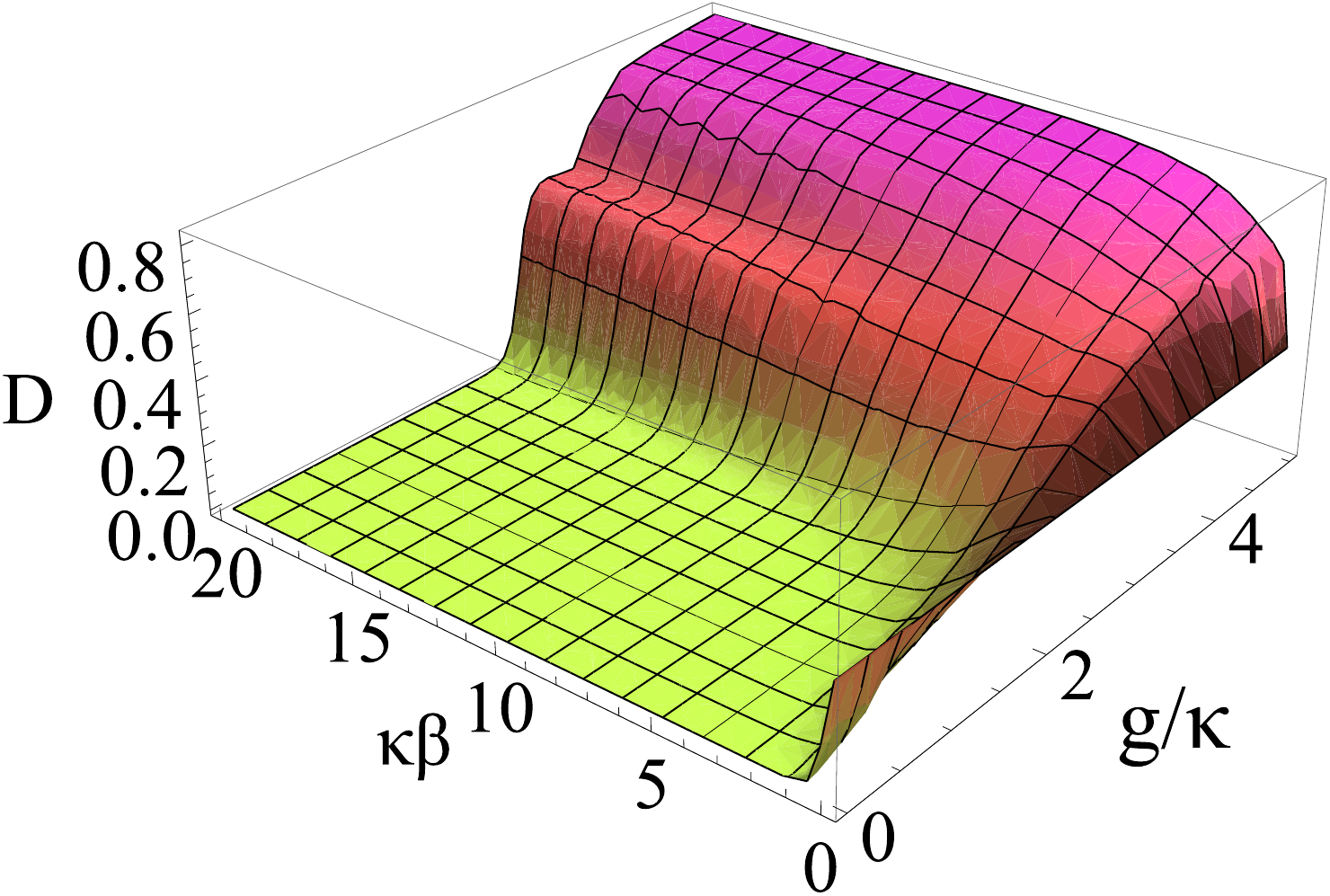}\\
\caption{(Color online) Trace distance between the Gibbs state and the product of its marginals at finite temperature is plotted as a function of the atom-field coupling strength $g$ and the inverse temperature $\beta$. At higher temperatures, the sudden change of value of the trace distance at the critical points is not so obvious due to thermal fluctuations.}\label{dst3d}
\end{figure}

\begin{figure}
\includegraphics[scale=.6]{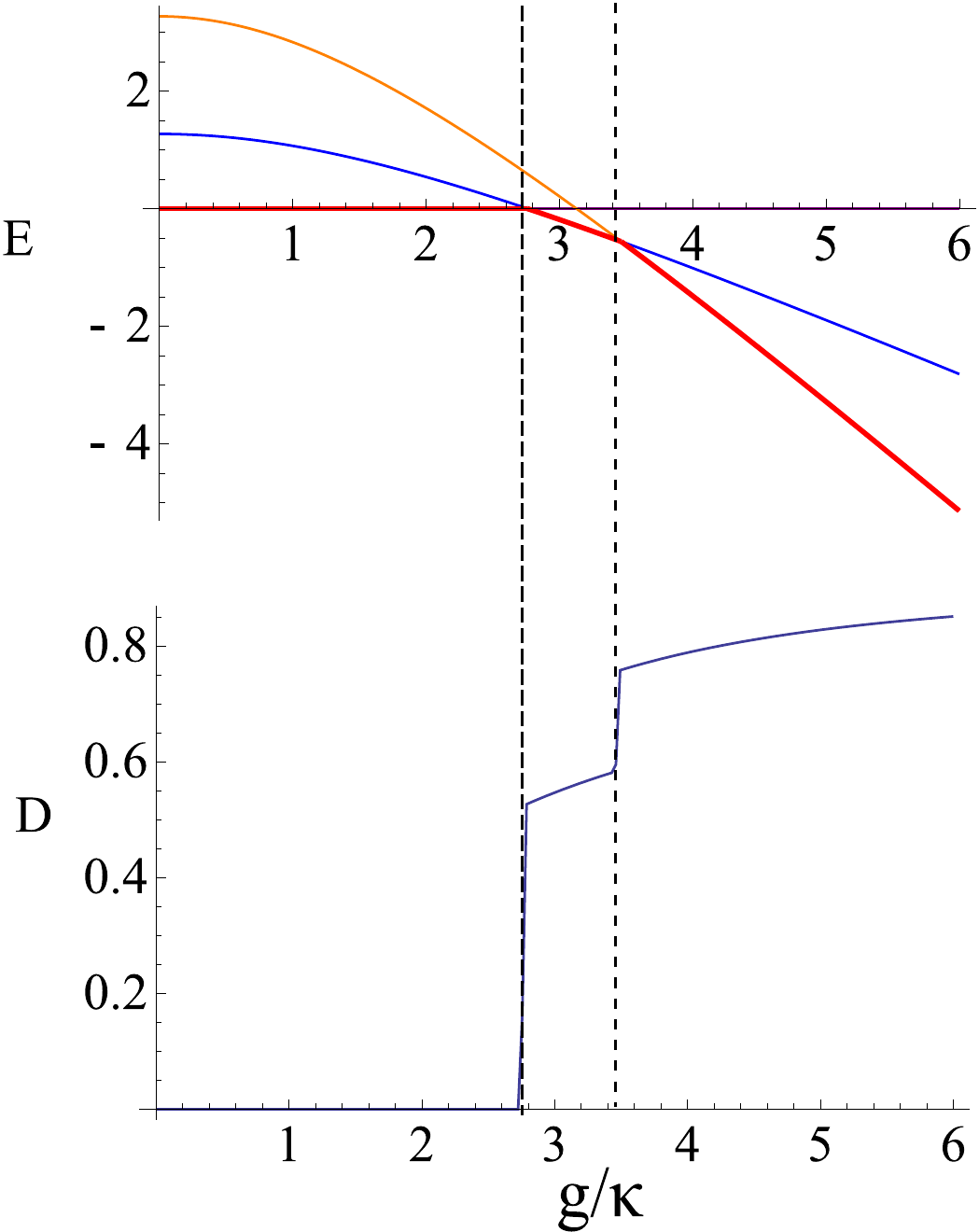}\\
\caption{(Color online) (Top) The three lowest-lying energy levels of the system as a function as a function of $g$, with the vertical black dashed line signifying the points where the ground state level crossing points, i.e. the phase transition points. (Bottom) Trace distance between the Gibbs state and the product of its marginals near zero temperature is plotted as a function of $g$ with $\kappa\beta=800$. All other parameters are taken to be the same as Fig.~\ref{dst3d}. At the critical points, the trace distance displays a sudden change in value.}\label{dst0}
\end{figure}

\begin{figure}
\includegraphics[scale=.5]{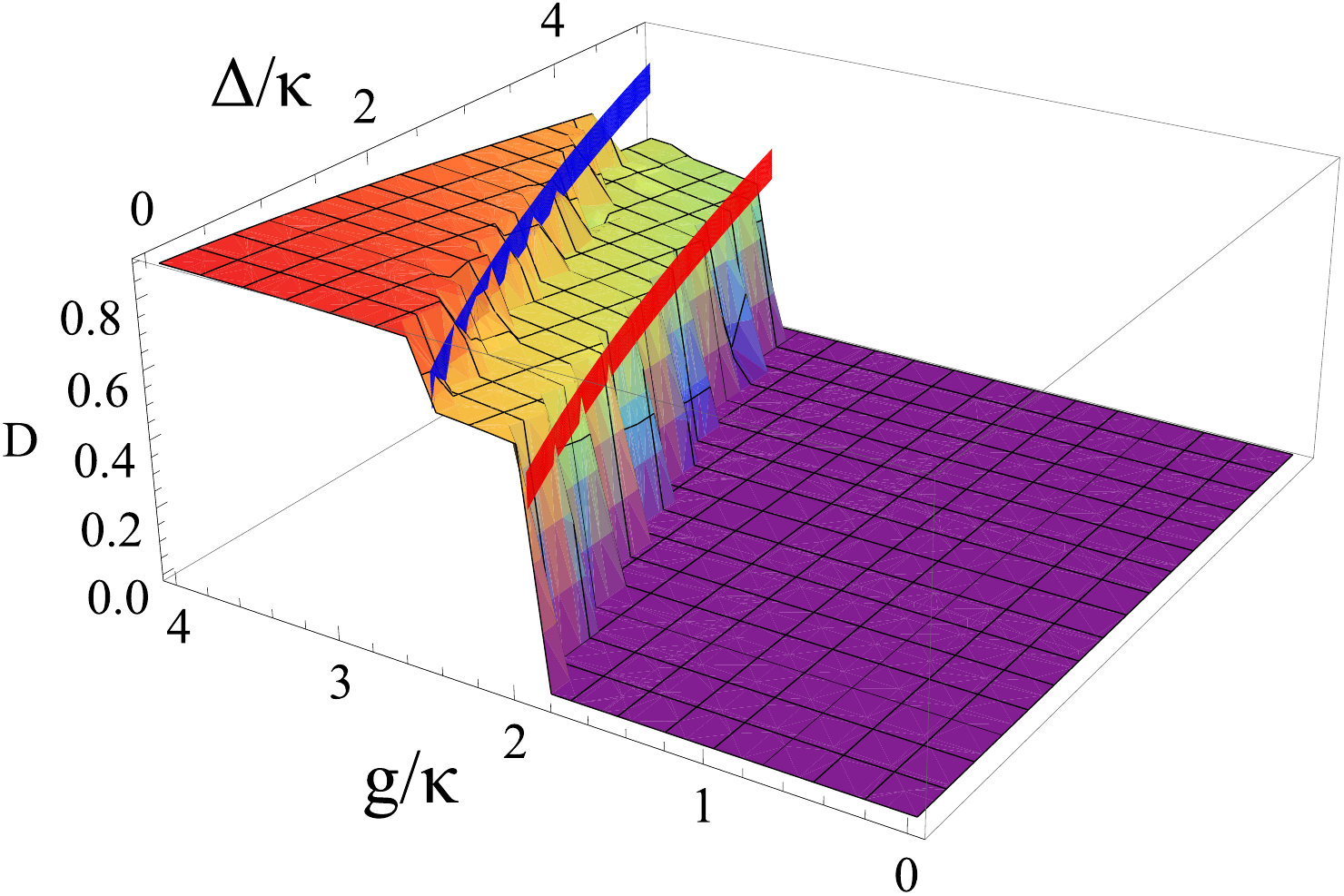}\\
\caption{(Color online) Trace distance between the Gibbs state and the product of its marginals at finite temperature is plotted as a function of the atom-field coupling strength $g$ and the detuning $\Delta$ near zero temperature $\kappa\beta=800$, and we have superimposed the critical points as the vertical blue and red surfaces. Along the critical points, the trace distance shows a sudden jump.}\label{dstph}
\end{figure}

We define the system to be all the atoms and the environment to be all field modes in our coupled Jaynes-Cumming lattice and consider the total initial thermal-equilibrium Gibbs state
\begin{equation*}
\rho_{SE}=e^{-\beta H}/Z
\end{equation*}
where the inverse temperature $\beta=1/k_b T$, $k_b$ the Boltzmann constant and $Z=\mathrm{Tr}(e^{-\beta H})$ is the partition function. We calculate the trace distance between the Gibbs state and the product of its marginal states $\rho_S\otimes\rho_E$. The reduced density matrix reads
\begin{align*}
\rho_S&=\mathrm{Tr}_E[\rho_{SE}]=\sum_{f_i}\langle f_i|\rho_{SE}|f_i\rangle,\\
\rho_E&=\mathrm{Tr}_S[\rho_{SE}]=\sum_{a_i}\langle a_i|\rho_{SE}|a_i\rangle,
\end{align*}
where $\mathrm{Tr}_E$ and $\mathrm{Tr}_S$ means the partial trace and is carried out by tracing over all field modes $|f_i\rangle$ to obtain the marginal for the atoms, and tracing over all possible combinations of atom states $|a_i\rangle$ to obtain the marginals for the field modes. It is noted that $\rho_{SE}$ and $\rho_S \otimes \rho_E$ have the same marginals for the system and environment, $\mathrm{Tr}_{E(S)}[\rho_{SE}] =\mathrm{Tr}_{E(S)}[\rho_S \otimes \rho_E]$, so the difference between the two density matrices measured by the trace distance can capture the system-bath correlation between the generic thermal Gibbs state of the global atoms plus fields system and the product of the marginals. With the full Hamiltonian diagonalized in the previous section, we can compute the trace distance for any number of excitons without the mean-field approximation which requires taking the thermodynamics limit. For states containing at most $N$ excitons, we first need to write down all the basis for the Hilbert space so that the matrix form of the Gibbs state density matrix along with the system and environment marginals can be obtained. Then, using Eq.~\eqref{dst_def0}, we can calculate the corresponding trace distance. For simplicity, we restrict ourselves to a physically rich space of maximal two excitons.  The basis of the space is chosen to be
\begin{align*}
\mathrm{I: } &|0,0\rangle_i^{\otimes N},\\
\mathrm{II: } &|0,1\rangle_i|0,0\rangle_j^{\otimes N-1},|1,0\rangle_i|0,0\rangle_j^{\otimes N-1},\\
\mathrm{III: } &|0,2\rangle_i|0,0\rangle_j^{\otimes N-1},|1,1\rangle_i|0,0\rangle_j^{\otimes N-1},\\
\mathrm{IV: } &|0,1\rangle_i|0,1\rangle_j|0,0\rangle_k^{\otimes N-2},|0,1\rangle_i|1,0\rangle_j|0,0\rangle_k^{\otimes N-2},\\
&|1,0\rangle_i|0,1\rangle_j|0,0\rangle_k^{\otimes N-2},|1,0\rangle_i|1,0\rangle_j|0,0\rangle_k^{\otimes N-2},i<j,
\end{align*}
where in each subspace spanned by each set of basis, the Hamiltonian is block diagonal, and the matrix elements of the Gibbs state in each set of basis is given by
\begin{align*}
\mathrm{I: } &1/Z&
\mathrm{II: } &\begin{bmatrix}x_i^{(1)}&z_i^{(1)}\\z_i^{(1)}&y_i^{(1)}\end{bmatrix}/Z\\
\mathrm{III: } &\begin{bmatrix}x_i^{(2)}&z_i^{(2)}\\z_i^{(2)}&y_i^{(2)}\end{bmatrix}/Z&
\mathrm{IV: } &M^{(ij)}/Z
\end{align*}
where
\begin{align*}
M^{(ij)}&=\begin{bmatrix}x_i^{(1)}&z_i^{(1)}\\z_i^{(1)}&y_i^{(1)}\end{bmatrix}\otimes \begin{bmatrix}x_j^{(1)}&z_j^{(1)}\\z_j^{(1)}&y_j^{(1)}\end{bmatrix}\\
x_i^{(n)}&=b_n(i)^2e^{-\beta E_n^+(i)}+a_n(i)^2e^{-\beta E_n^-(i)}\\
y_i^{(n)}&=a_n(i)^2e^{-\beta E_n^+(i)}+b_n(i)^2e^{-\beta E_n^-(i)}\\
z_i^{(n)}&=a_n(i)b_n(i)\left[e^{-\beta E_n^+(i)}-e^{-\beta E_n^-(i)}\right].
\end{align*}
The partition function is, therefore,
\begin{align}
Z&=1+\sum_i\left(x_i^{(1)}+y_i^{(1)}+x_i^{(2)}+y_i^{(2)}\right)+\sum_{i<j}\mathrm{Tr}[M^{(ij)}]
\end{align}
We plot the trace distance $D(\rho_{SE},\rho_S\otimes\rho_E)$ at finite temperature as a function of the atom-field coupling strength $g$ and the inverse temperature $\beta$ in Fig.~\ref{dst3d}. We can see from Fig.~\ref{dst3d} that the trace distance shows a sudden jump for increasing coupling strength $g$ at the critical points at finite temperatures. For finite low temperatures, as the coupling strength $g$ increases, the condition in Eq.~\eqref{gnd_cr_g} is met, a quantum phase transition takes place and the ground state is dramatically different, resulting in a sudden change of value in the trace distance, which means that the trace distance can be used to locate the critical points of QPT at finite temperatures. As the temperature approaches absolute zero, the discontinuity of the trace distance becomes more pronounced, and at higher temperatures, this discontinuity is not so obvious due to thermal fluctuations. Taking $\kappa\beta=800$, we plot the trace distance $D(\rho_{SE},\rho_S\otimes\rho_E)$ in Fig.~\ref{dst0} with the energy spectrum of the three lowest-lying energy levels as a function of the atom-field coupling strength $g$. The vertical black dashed lines signifies the points where the ground state level crossing takes place, and the QPT takes place. As expected, when that happens, the lowest-lying eigen-energy takes a different form, and there is a sudden jump of the trace distance. The trace distance $D(\rho_{SE},\rho_S\otimes\rho_E)$ at near-zero temperature as a function of the atom-field coupling strength $g$ and the detuning $\Delta$ is plotted in Fig.~\ref{dstph} with the critical points superimposed. It is clear that the trace distance changes suddenly across different phases, and is dependent on both the atom-field coupling strength $g$ and the detuning $\Delta$ according to Eq.~\eqref{gnd_cr_g}, and as the detuning gets larger, the QPT only happens for larger atom-field coupling strengths.

\begin{figure}
\includegraphics[scale=.5]{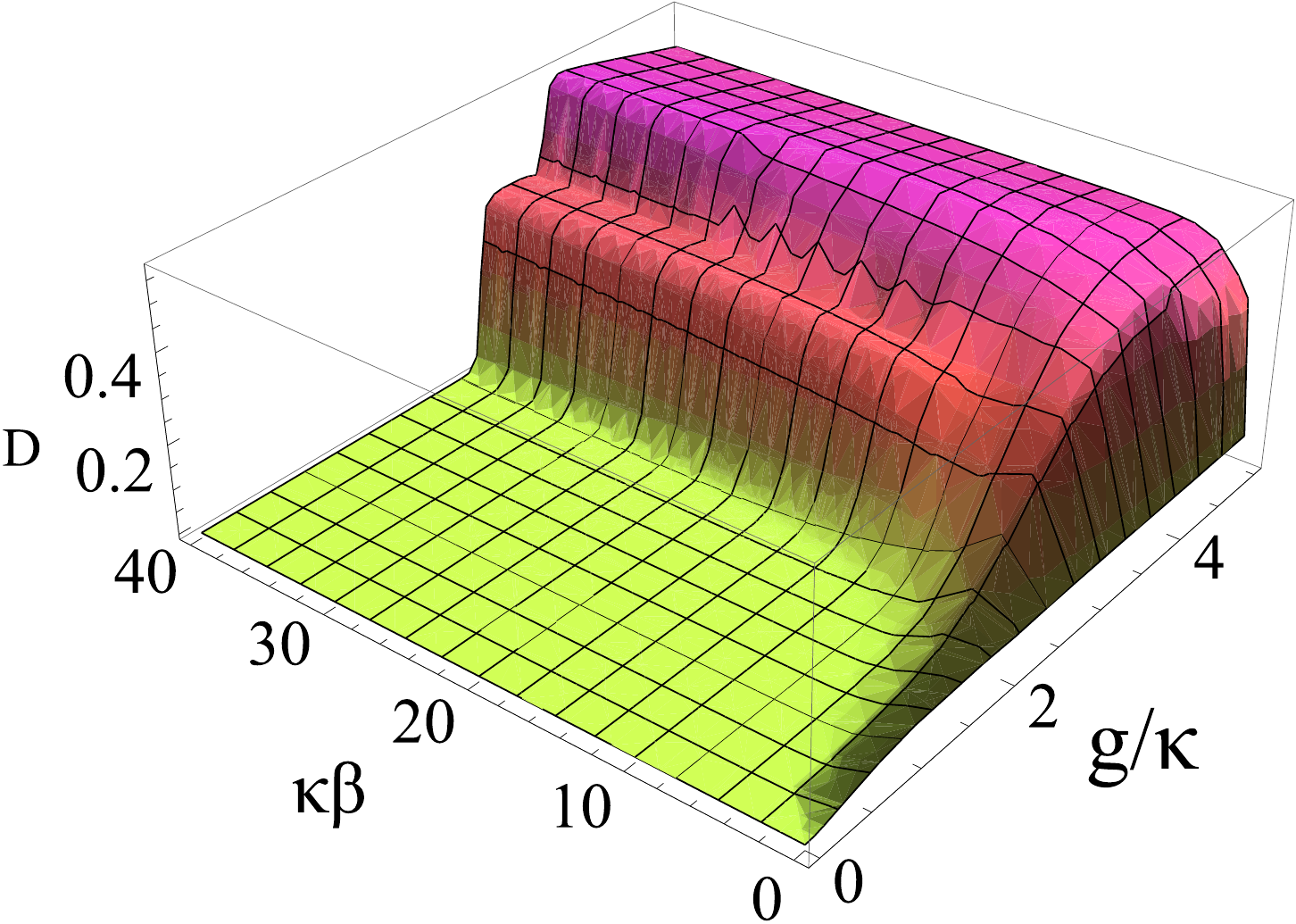}\\
\caption{(Color online) The maximum value of trace distance $D(\rho_s(0),\rho_s(t))$ is plotted as a function of the atom-field coupling strength $g$ and the inverse temperature $\beta$ where $\rho_s$ is the atom part of the Gibbs state. At higher temperatures, the sudden change of value of the trace distance at the critical points is not so obvious due to thermal fluctuations.}\label{dstt3d}
\end{figure}

\begin{figure}
\includegraphics[scale=.6]{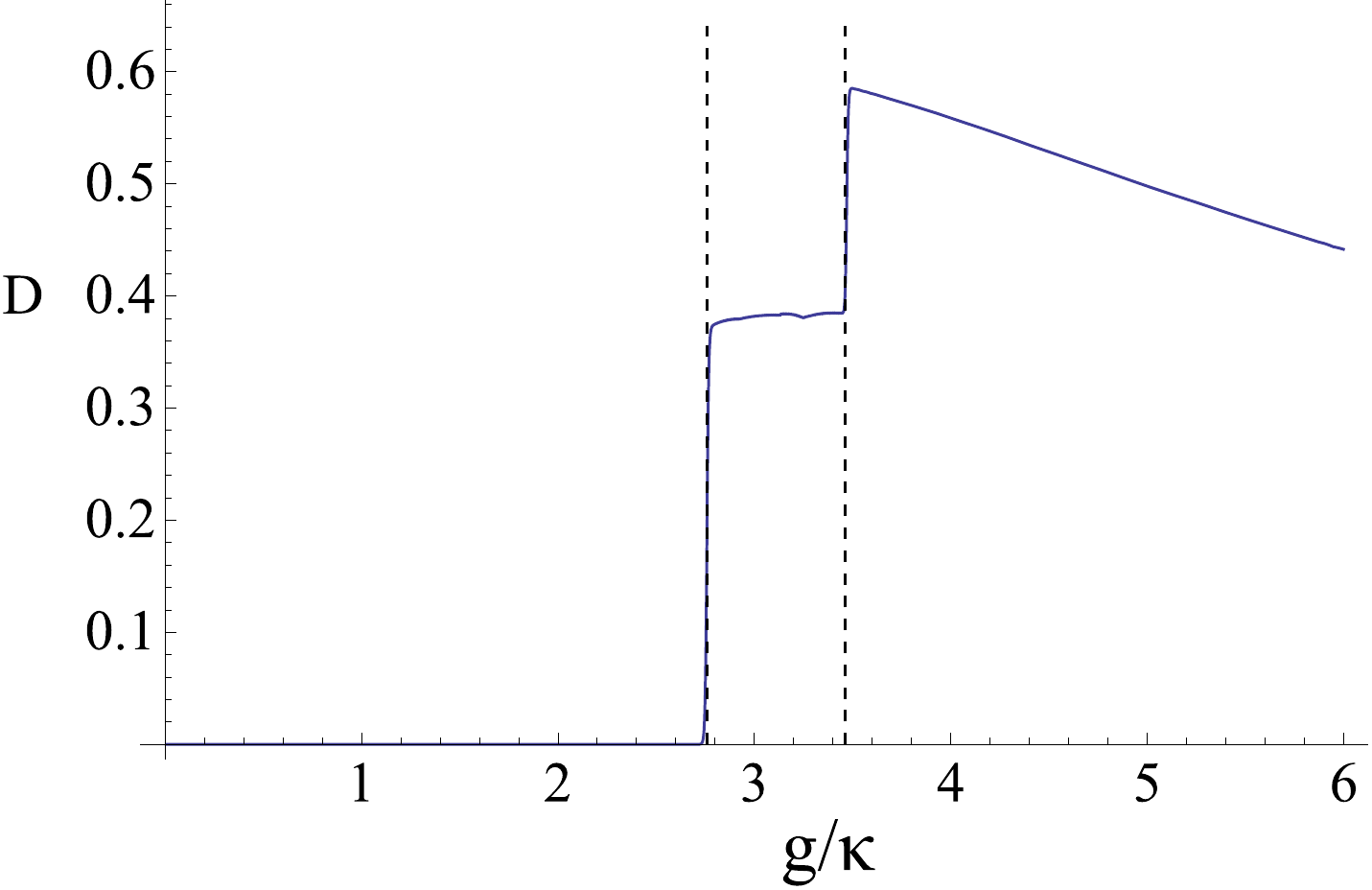}\\
\caption{(Color online) The maximum value of trace distance $D(\rho_s(0),\rho_s(t))$ is plotted as a function of the atom-field coupling strength $g$ at the inverse temperature $\kappa\beta=300$. All other parameters are taken to be the same as Fig.~\ref{dstt3d} and the vertical dashed line corresponds to the critical points.}\label{dstt0}
\end{figure}

We now consider the time evolution of trace distance and its relationship with QPT for non-equilibrium states. Since all trace-preserving positive maps $\Lambda$ are contractions of the trace distance~\cite{Smirne2010}
\begin{equation*}
D(\Lambda\rho_1,\Lambda\rho_2)\leq D(\rho_1,\rho_2),
\end{equation*}
it should be possible to detect QPT with the upper bound of time evolution of trace distance. Because the Gibbs state is in thermal equilibrium, the state does not evolve in time. Therefore, we take the initial state as the product of its marginals and calculate the trace distance of the reduced density matrix of the atom part at time $0$ and at time $t$. With the diagonalized Hamiltonian, the time evolution problem can be easily solved. We plot the maximum value of trace distance $D(\rho_{S}(0),\rho_S(t))$ at finite temperature as a function of the atom-field coupling strength $g$ and the inverse temperature $\beta$ in Fig.~\ref{dstt3d}. As depicted, the maximum value of trace distance also shows a sudden jump at the critical points, which means the trace distance can detect QPT for non-equilibrium cases. Taking $\kappa\beta=300$, we also plot the trace distance $D(\rho_{S}(0),\rho_S(t))$ in Fig.~\ref{dstt0}, which still agrees with the result obtained using Gibbs states.

\begin{figure}
  \includegraphics[scale=.6]{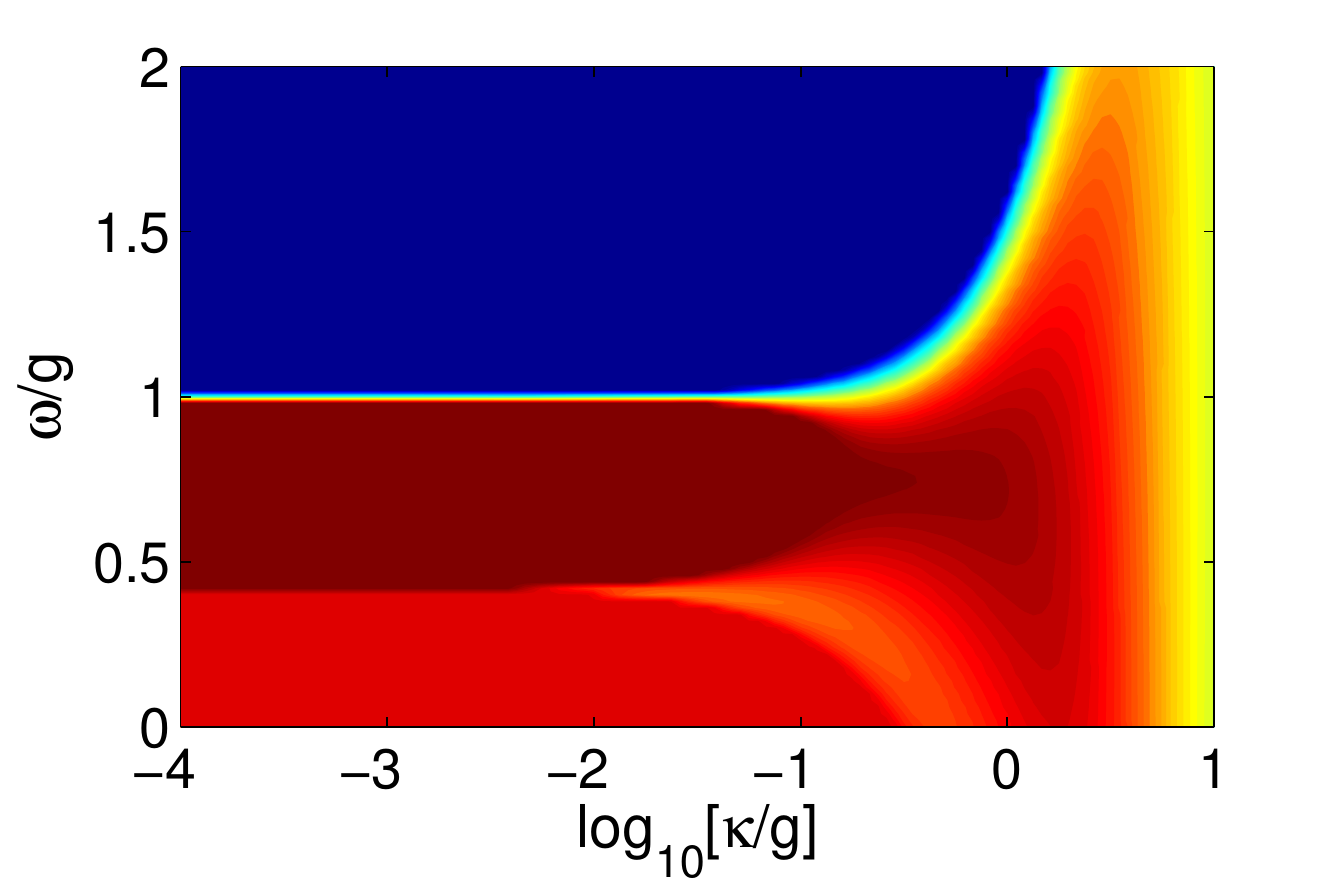}\\
  \caption{(Color online) Contour plot of the trace distance between the Gibbs state and the product of its marginals at zero detuning. It can be readily seen that the trace distance can indicate the critical points of QPT in the mean-field analysis.}\label{mfld}
\end{figure}

In order to compare our results with that of the mean-field theory, we now calculate the trace distance for Gibbs states, where the decoupling approximation $b_i^\dagger b_j=\langle b_i^\dagger\rangle b_j+\langle b_j\rangle b_i^\dagger-\langle b_i^\dagger\rangle\langle b_j\rangle$. The corresponding trance distance is displayed in Fig.~\ref{mfld}. It can be seen that the QPT boundary obtained via trace distance agrees well with that obtained using the usual mean-field approach and the one-polarization approximation~\cite{Greentree2006,Houck2012,Quach2009}.

\section{Scaling behavior of the JCH lattice at critical points}\label{sec_scal}

\begin{figure}
\includegraphics[scale=.6]{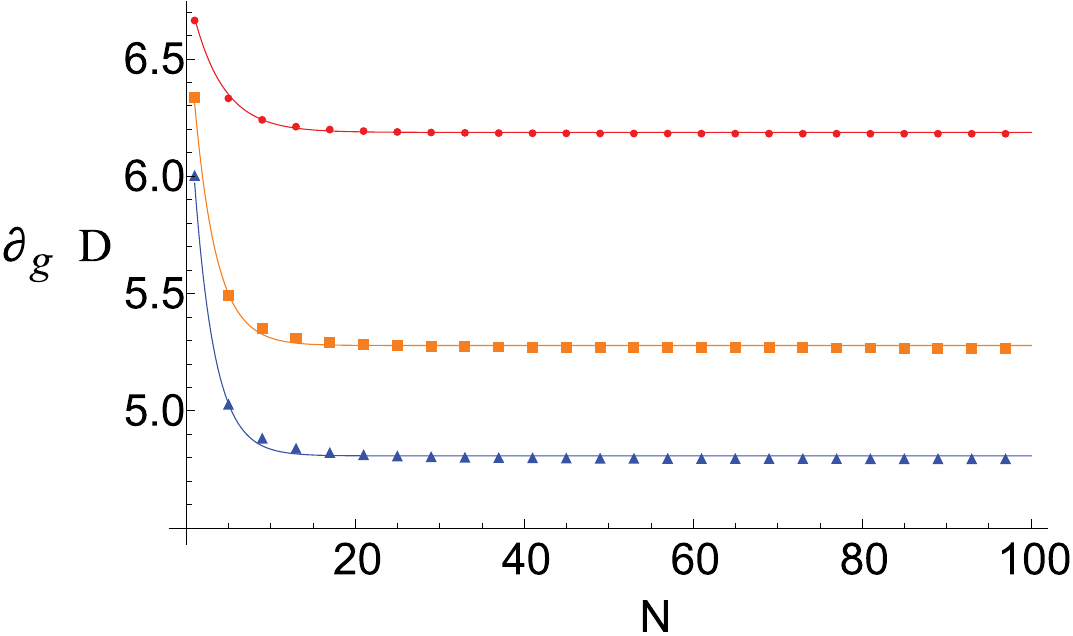}\\
\caption{(Color online) Scaling behavior of the first derivative of the trace distance between the Gibbs state and the product of its marginals at the critical points at near zero temperature $\kappa\beta=300$. The circles(red), squares(orange) and triangle(blue) marks correspond to $\Delta_f/\kappa=0,3,5$ respectively and are obtained analytically. The curves are obtained using least-square fit of the form $f(N)=Ae^{-bN}+C$.}\label{dstsc}
\end{figure}

The scaling behavior at critical points is a very important part of the study of QPTs~\cite{Sachdev1999}. Since the trace distance $D(\rho_{SE},\rho_S\otimes\rho_E)$ can be considered as a form of system-environment correlation~\cite{Smirne2010}, and motivated by the study of QPT in spin chain systems~\cite{Sachdev1999,Osborne2002}, we study the derivative of the trace distance against the atom-field coupling strength $g$. Because we are only interested in the behavior of the trace distance at the first critical point, we can safely discard the two-exciton subspace. The Hamiltonian and Gibbs state is still block diagonal, and the trace distance is given by
\begin{align}
&D(\rho_{SE},\rho_S\otimes\rho_E)=\frac{1}{2}|\frac{(1+\sum_i x_i^{(1)})(1+\sum_i y_i^{(1)})}{Z'}-\frac{1}{Z}|\nonumber\\
&+\frac{1}{4}\sum_i\left(|\rho x_i+\rho y_i-\sqrt{(\rho x_i-\rho y_i)^2+4\rho z_i^2}|\right.\nonumber\\
&\left.+|\rho x_i+\rho y_i+\sqrt{(\rho x_i-\rho y_i)^2+4\rho z_i^2}|\right),\label{dpg}
\end{align}
where
\begin{align*}
Z&=1+\sum_i\left(x_i^{(1)}+y_i^{(1)}\right),\\
Z'&=(1+\sum_i x_i^{(1)})(1+\sum_i y_i^{(1)})+(1+\sum_i x_i^{(1)})\sum_i x_i^{(1)}\\
&+(1+\sum_i y_i^{(1)})\sum_i y_i^{(1)},\\
\end{align*}
and
\begin{align*}
\rho x_i&=\frac{(1+\sum_i x_i^{(1)})x_i^{(1)}}{Z'}-\frac{x_i^{(1)}}{Z},\\
\rho y_i&=\frac{(1+\sum_i y_i^{(1)})y_i^{(1)}}{Z'}-\frac{y_i^{(1)}}{Z},\\
\rho z_i&=-\frac{z_i^{(1)}}{Z},
\end{align*}
from which the derivative $\partial_g D(\rho_{SE},\rho_S\otimes\rho_E)$ can be readily calculated. We plot the value of $\partial_g D(\rho_{SE},\rho_S\otimes\rho_E)$ at the critical points for JCH array with size $N=1\ldots 100$ with $\Delta_f/\kappa=0,3,5$ in Fig.~\ref{dstsc}. Least-square fit is used for the derivative for JCH with size $N$ as $f(N)=Ae^{-bN}+C$, and we find that the derivative has a exponential scaling behavior as the system size grows. We can see from Fig.~\ref{dstsc} the fitted curve agrees well with the values obtained from Eq.~\eqref{dpg}, and different detuning leads to different scaling behaviors, and smaller detuning leads to bigger derivatives, meaning the trace distance changes more rapidly for smaller detunings.

\section{Conclusions}\label{sec_con}

We investigated how to use trace distance to detect the critical points of QPT in a JCH lattice at finite temperature. It is found that the trace distance shows a sudden jump at the phase transition points at low temperatures which means that that the trace distance can be used to describe the critical points of QPT. The critical points are found to be dependent on the atom-field interaction strength $g$ and the detuning factor $\Delta$. For non-equilibrium states, we show that the time evolution of the trace distance's maximum value is also a good indicator of the critical points. Our results agree well with mean field analysis. Finally, the scaling behavior of the derivative of the trace distance is found to exist at the critical points, and the scaling rule is dependent on the system parameters. Traditional QPT approaches mainly focus on the identification of the order parameters and the pattern of symmetry breaking. The trace distance approach presented in this paper allows us to detect QPTs without any prior knowledge of order parameters and may be extended to other many-body systems.

\section{Acknowledgments}

This project was supported by the National Natural Science Foundation of China (Grant No. 11274274).

%%%%%%%%%%%%%%%%%%%%%%%%%%%%%%%%%%%%%%%%%%%%%%%%%%%%%%%%%%%%%%%%%%%%%%%%%%%%

%%%%%%%%%%%%%%%%%%%%%%%%%%%%%%%%%%%%%%%%%%%%%%%%%%%%%%%%%%%%%%%%%%%%%%%%%%%%

\end{document}